\documentclass[doublecol]{epl2}
\usepackage{amsmath}
\usepackage{amssymb}

\def\be{\begin{equation}}
\def\ee{\end{equation}}
\def\figsize{8.5truecm}

\title{Hyperscaling violation
in the 2D 8-state Potts model
with long-range correlated disorder}

\author{C. Chatelain\inst{1,2}}
\institute{
\inst{1}
School of Physics, Indian Institute of Science
Education and Research (IISER),
Thiruvananthapuram, India \\
\inst{2}
Groupe de Physique Statistique,
D\'epartement P2M,
Institut Jean Lamour (CNRS UMR 7198),
Universit\'e de Lorraine, France
}

\abstract{
The first-order phase transition of the two-dimensional eight-state
Potts model is shown to be rounded when long-range correlated
disorder is coupled to energy density. Critical exponents are
estimated by means of large-scale Monte Carlo simulations.
In contrast to uncorrelated disorder, a violation of the
hyperscaling relation $\gamma/\nu=d-2x_\sigma$ is observed.
Even though the system is not frustrated, disorder fluctuations
are strong enough to cause this violation in the
very same way as in the 3D random-field Ising model.
In the thermal sector too, {evidence is} given for such
violation in the two hyperscaling relations
$\alpha/\nu=d-2x_\varepsilon$ and $1/\nu=d-x_\varepsilon$.
In contrast to the random field Ising model, at least two
hyperscaling violation exponents are needed.
The scaling dimension of energy is conjectured to be
$x_\varepsilon=a/2$, where $a$ is the exponent of
the algebraic decay of disorder correlations.
}

\pacs{64.60.De}{Statistical mechanics of phase transitions in model systems}
\pacs{05.50.+q}{Potts models in lattice theory and statistics}
\pacs{05.70.Jk}{Critical phenomena in thermodynamics}
\pacs{05.10.Ln}{Monte Carlo methods in statistical physics and nonlinear dynamics}
\begin{document}
\maketitle

\section{Introduction}
Quenched disorder when coupled to the energy density, say by dilution or random
couplings, is known to soften first-order phase transitions. As argued
by Imry and Wortis~\cite{ImryWortis}, local fluctuations of
impurity concentration can destabilise the ordered phases in coexistence at
the transition temperature if the surface tension is sufficiently small.
%, each cluster will undergo a transition into
%the disordered phase at a different temperature. The jump of the total
%magnetisation will be soften and may even be completly rounded.
%Experimental observations of such a complete smoothing
%of a first-order phase transition are numerous,
%the most famous one being the nematic-isotropic transition of the
%$4'-n-{\rm octyl}-4-{\rm cyanobiphenyl}$
%8CB liquid crystal~\cite{Wu}.
In 2D, it was rigorously proved that an infinitesimal amount of disorder
is sufficient to make any first-order transition continuous~\cite{HuiBerker,
Aizenman}. The complete vanishing of the latent heat was first observed
numerically in the case of the 2D 8-state Potts model~\cite{Landau}.
The critical behaviour of the disorder-induced second-order phase
transition is governed by a new random fixed point~\cite{Chatelain}.
The universality class was later shown to
depend on the number of states $q$~\cite{RBPotts}.
In 3D, a finite disorder is required to round completely the
first-order phase transition. The phase diagram exhibits
a tricritical point separating a first-order regime from the
disorder-induced continuous one, as first observed in the bond-diluted
4-state Potts model~\cite{Chatelain3D}.
%Again, the second regime is
%governed by a unique random fixed point, though with strong corrections
%to scaling.
A rounding of the first-order phase transition of the 2D Potts
model was also reported for anisotropic aperiodic sequences of
couplings~\cite{PEBerche}, and for layered random couplings~\cite{Senthil,
Carlon}. In both cases, the couplings are infinitely correlated in one
direction. In the random case, the critical behaviour was shown to be
governed by a $q$-independent infinite-randomness fixed point.
The critical exponents are therefore those of the layered random
Ising model, the celebrated McCoy-Wu model~\cite{McCoyWu,DFisher95}.
Interestingly, the same critical behaviour is observed for the
Potts model with homogeneous uncorrelated disorder in the limit
$q\rightarrow +\infty$~\cite{Mercaldo}.

%Note finally that the above-mentionned models with infinitely
%correlated disorder in one direction can be mapped onto random
%quantum systems. Hui and Berker's arguments for the rounding of
%a classical first-order phase transition have been extended to
%the quantum case~\cite{Goswami}, as well as Aizenmann and Wehr
%rigourous proof~\cite{Lebowitz}.

In this letter, we consider the case of random bond couplings
$J_{ij}>0$ with algebraically decaying correlations
$\overline{(J(0)-\bar J)(J(\vec r)-\bar J)}\sim r^{-a}$.
{
According to the Imry-Wortis criterion, the low-temperature
phase is destabilised when the fluctuations of exchange energy
inside a ferromagnetic domain of characteristic length $\ell$,
increase faster with $\ell$ than the interface free energy
$\sigma\ell^{d-1}$. Since the contribution of correlations to
these fluctuations reads
     \be\sqrt{\overline{\big[\sum_{i,j}(J_{ij}-\bar J)\big]^2}}
     \sim\left[\ell^d\int_{\ell^d}
     {d^d\vec r\over r^a}\right]^{1/2}\sim \ell^{d-a/2},\quad (a\le d)\ee
we expect the first-order phase transition to be softened
for $a\le 2$ in the two-dimensional case. For $a>d$ and uncorrelated
disorder, the main contribution is due to the fluctuation term
$\sqrt{\sum_{i,j}\overline{(J_{ij}-\bar J)^2}}\sim \ell^{d/2}$.
The case of an uncorrelated disorder is therefore equivalent to $a=d$.}
Unexpectedly, we observe that the critical exponents at the
randomness-induced second-order phase transition do not satisfy
hyperscaling relations. Such a violation had only been reported
in random systems with frustration, spin glasses or random-field
systems, but, to our knowledge, never for purely ferromagnetic
systems. In the first section of this letter, the details of the
Monte Carlo simulation are presented. Hyperscaling violation is
studied first in the magnetic sector and then in the energy
sector.

\section{Description of the simulation}
We consider the 2D $q$-state Potts model with Hamiltonian
   \be-\beta H=\sum_{(i,j)} J_{ij}\delta_{\sigma_i,\sigma_j}\ee
where $\sigma_i\in\{0,1,\ldots,q-1\}$ and the sum extends over pairs
of nearest neighbours of the square lattice.
We restrict ourselves to the case $q=8$ for which the correlation length
of the pure model is $\xi\simeq 24$ at the transition temperature.
{The order parameter $m$ is defined as
     \be m={q\rho_{\rm max}-1\over q-1},\quad
     \rho_{\rm max}={\rm max}_{\sigma} {1\over N}\sum_i
     \delta_{\sigma_i,\sigma}\ee
where $\rho_{\rm max}$ is the density of spins in the majority state.
This definition breaks the ${\mathbb Z_q}$ symmetry of the Hamiltonian
in the same way as an infinitesimal magnetic field. In the case $q=2$
corresponding to the Ising model, the usual order parameter
${1\over N}\langle |\sum_i\sigma_i|\rangle$ is recovered.}

We consider a binary distribution of coupling constants $J_{ij}
\in\{J_1,J_2\}$ with
    \be\big(e^{J_1}-1\big)\big(e^{J_2}-1\big)=q.\label{SelfDual}\ee
In the case of uncorrelated disorder, Eq. (\ref{SelfDual}) is the
self-duality condition that gives the location of the critical line.
The ratio $r=J_2/J_1$ is used as a measure of the strength of disorder.
Here, we present results for the case $r=8$.
To generate correlated coupling configurations $\{J_{ij}\}$,
we simulate another spin model, namely the Ashkin-Teller model
($\sigma_i,\tau_i=\pm 1$)
   \be -\beta H^{\rm AT}=\sum_{(i,j)} \big[J^{\rm AT}\sigma_i\sigma_j
   +J^{\rm AT}\tau_i\tau_j+K^{\rm AT}\sigma_i\sigma_j\tau_i\tau_j\big]\ee
at different points of its critical line $e^{-2K^{\rm AT}}=\sinh 2J^{\rm AT}$.
Two symmetries of the Hamiltonian are spontaneously broken at low temperatures:
the global reversal of the spins $\sigma_i$ and the reversal of both
$\sigma_i$ and $\tau_i$. Therefore, two order parameters can be defined,
magnetisation $\sum_i\sigma_i$ and polarisation $\sum_i \sigma_i\tau_i$,
leading to two independent scaling dimensions:
    \be\beta_\sigma^{\rm AT}={2-y\over 24-16y},\quad\quad
    \beta_{\sigma\tau}^{\rm AT}={1\over 12-8y}\ee
wherein we use the parametrization
    \be\cos{\pi y\over 2}={1\over 2}\left[e^{4K^{\rm AT}}-1\right],
    \quad (y\in[0;4/3]).\ee
The correlation length exponent is $\nu^{\rm AT}={2-y\over 3-2y}$.
Spin configurations of this model are generated by Monte Carlo simulation
using a cluster algorithm~\cite{Salas}. For each of them, a coupling
configuration of the Potts model is constructed as
    \be J_{ij}={J_1+J_2\over 2}+{J_1-J_2\over 2}\sigma_i\tau_i,\ee
where the site $j$ is either at the right or below the site $i$.
{This construction ensures that the constraint (\ref{SelfDual})
implies the self-duality of our random Potts model. }
%Because $\langle\sigma\tau\rangle=0$ on the critical line, we expect
%$\bar J={J_1+J_2\over 2}$.
{
Since the Askin-Teller is considered on its critical line,
} disorder fluctuations are self-similar and
the coupling constants display algebraic correlations
    \be\overline{(J_{ij}-\bar J)(J_{kl}-\bar J)}
    \sim |\vec r_i-\vec r_k|^{-a} \label{CorrAT}\ee
at large distances with
{
   \be a=2\beta_{\sigma\tau}^{\rm AT}/\nu^{\rm AT}={1\over 4-2y}.\ee
} We have considered six points on the critical line,
$y\in\{0,0.25,0.50,0.75,1,1.25\}$,
leading to six correlated disorder distributions with
$a\simeq 0.25$, $0.286$, $0.333$, $0.4$, $0.5$ and $0.667$.
{
We have checked that disorder fluctuations display the expected
scaling behaviour. The electric susceptibility of the Ashkin-Teller
model, defined as
    \be\chi^{\rm AT}=L^d\big[\overline{p^2}
    -\overline{{|p|}}^2\big]\label{ChiAT}\ee
where $p={1\over N}\sum_i \sigma_i\tau_i$ is the polarisation density,
is computed. According to the fluctuation-dissipation theorem, this
quantity is equal to the integral of the correlations $(\ref{CorrAT})$
over the volume of the system. It is therefore expected to scale as
$L^{d-a}$. The data are plotted on figure~\ref{Corr-AT}. A nice power-law
behaviour is observed over the whole range of lattice sizes that are
considered. The {fitted} exponents are given in Tab.~\ref{Tab2}
for all values of $y$. For $y\ge 0.75$, they are compatible with exact
exponents. For smaller values of $y$, the deviation to the exact result
is at most of $5\%$. Note that for $y=0$, the Ashkin-Teller model is
equivalent to the 4-state Potts model. The critical behaviour is
therefore affected by logarithmic corrections.
These values have been obtained with, and only with,
the coupling configurations used during the simulation of
the Potts model with correlated disorder. The agreement with
the expected values, even for small lattice sizes, indicates
that the number of disorder configurations is sufficient
to reproduce correctly the expected disorder fluctuations.
}

{
For comparison, simulations for the Potts model with uncorrelated
disorder are also performed. The same simulation code is used
but with an infinite temperature of the Ashkin-Teller model in
order to obtain uncorrelated spins and therefore uncorrelated
couplings $J_{ij}$.
}

{
\begin{figure}
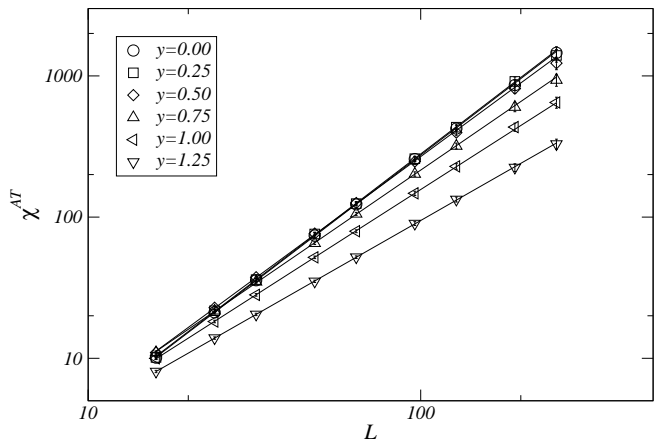

\onefigure[width=\figsize]{CorrAT.eps}
\caption{{Electric susceptibility $\chi^{\rm AT}$ of the Ashkin-Teller
model, or equivalently fluctuations of the couplings $J_{ij}$, versus
the lattice size $L$. The different symbols correspond to different
values of the parameter $y$ of the Ashkin-Teller model. The straight lines
are power-law {fits} to the data.}}
\label{Corr-AT}
\end{figure}
}

{
\begin{largetable}
\caption{{Critical exponents $d-a$ of the electric susceptibility
$\chi^{\rm AT}$ of the Ashkin-Teller model, or equivalently of fluctuations
of the couplings $J_{ij}$, for different values of the parameter $y$. The
numerical estimate of $d-a$ is denoted by MC (second line) and the exact
value by Th. (third line).}}
\label{Tab2}
\begin{center}
\begin{tabular}{|l|llllll|}
\hline
$y$ & $0$ & $0.25$ & $0.5$ & $0.75$ & $1$ & $1.25$ \\
%\hline
%$a$ & $0.25$ & $0.286$ & $0.333$ & $0.4$ & $0.5$ & $0.667$ \\
\hline
$d-a$ (MC) & $1.80(3)$ & $1.80(3)$ & $1.73(4)$ & $1.61(3)$ & $1.50(3)$
& $1.34(2)$\\
$d-a$ (Th.) & $1.75$ & $1.714$ & $1.667$ & $1.600$ & $1.500$ & $1.333$ \\
\hline
\end{tabular}
\end{center}
\end{largetable}
}

The Potts model is then simulated using the Swendsen-Wang algorithm~\cite{SW}.
Lattice sizes between $L=16$ and 256 are considered. For each disorder
configuration, $1000$ MCs are performed to thermalise the system
and $20,000$ MCs for data accumulation (auto-correlation time is
$\tau\simeq 2$ for $L=256$). Thermodynamic quantities
are averaged over a number of disorder configurations proportional
to $1/L^2$. For the largest lattice size $L=256$, 2560 disorder
configurations are generated while for $L=64$ for instance, this
number is raised up to $40960$. Stability of disorder averages is
checked. In the following, we will denote $\langle X\rangle$ the average
of an observable over thermal fluctuations and $\overline{\langle X\rangle}$
the average of the latter over disorder.

On the critical line, the typical spin configurations of the
Ashkin-Teller model display a large cluster of polarisation $\sigma\tau=+1$
or $-1$. As a consequence, our random coupling configurations
also exhibit large clusters of either strong or weak bonds. The
probability distribution of the total energy of the Potts model
shows two peaks corresponding to these two kinds of bond configurations.
This distribution is highly correlated to the probability distribution
of polarisation of the Ashkin-Teller model. Since the latter undergoes
a second-order phase transition, the two peaks come closer
as the lattice size is increased. Note that in our model, macroscopic
region of strong couplings are not rare: they have a probability $1/2$.
Moreover, they have a fractal dimension $1<d_f<2$ determined by the
Ashkin-Teller model and, in the thermodynamic limit, only one such
macroscopic region is expected to be present in the system. For this
reason, the transition is not smeared but rounded~\cite{Vojta}.

\section{Magnetic sector}
\begin{figure}
\onefigure[width=\figsize]{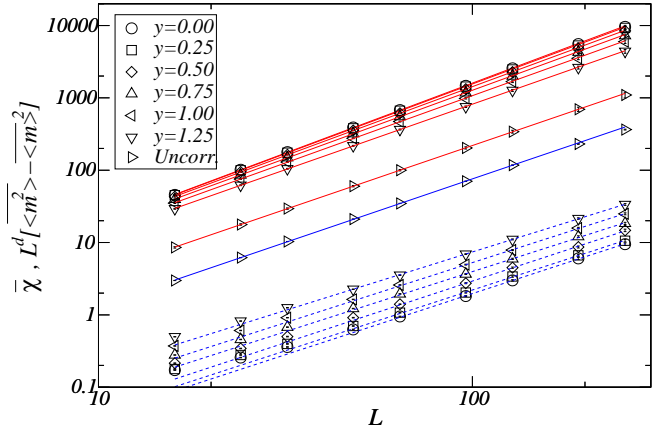}
\caption{Average magnetic susceptibility $\bar\chi$ (bottom)
and $L^d[\overline{\langle m^2\rangle}-\overline{\langle m\rangle}^2]$
(top) versus lattice size $L$ for different values of $y$ and
for uncorrelated disorder (Uncorr.). The straight lines
correspond to power-law {fits}. {Dashed lines indicate that
the fit} was performed over lattice sizes $L\ge 96$ only.}
\label{Chi}
\end{figure}

\begin{figure}
\onefigure[width=\figsize]{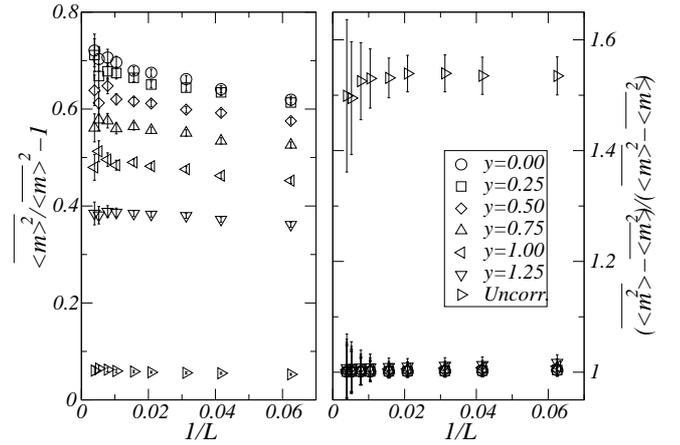}
\caption{{
On the left, ratio defined by Eq. (\ref{Rm}) for different
values of $y$ and for uncorrelated disorder (Uncorr.) versus the
inverse of the lattice size. The non-vanishing asymptotic limit, i.e.
the extrapolated value on the $y$-axis, indicates that magnetization
is not self-averaging for both correlated and uncorrelated disorder.
On the right, ratio (\ref{RatioChi}) of the two terms whose difference
gives the average susceptibility. Asymptotically, this ratio displays
a plateau at a value compatible with 1 for correlated disorder. For
uncorrelated disorder, the asymptotic value is incompatible with $1$.
}}
\label{RM}
\end{figure}

We estimate critical exponents by Finite-Size Scaling.
The exponent $\beta/\nu$ can be extracted from magnetisation
$\overline{\langle m\rangle}$ and its moments
$\overline{\langle m^n\rangle}$ with $n=2,3,4$. We observe
nice power laws without any significant correction to scaling.
Our estimates of $x_\sigma=\beta/\nu$ evolve with $a$ and range
from $0.061(5)$ ($y=0$) to $0.108(4)$ ($y=1.25$), to be compared
with $0.150(2)$ for uncorrelated disorder (Tab.~\ref{Tab1}).
We then consider the average magnetic susceptibility, numerically
computed via the fluctuation-dissipation theorem
  \be\bar\chi=L^d\overline{\langle m^2\rangle-\langle m\rangle^2}.\ee
The data display large corrections to scaling (see Fig.~\ref{Chi}).
A cross-over is observed around $L=48$, not far from the correlation
length $\xi=24$ of the pure 8-state Potts model. In the region
$L\ge 96$, power-law {fits} give stable estimates for
$\gamma/\nu$ going from $1.70(7)$ ($y=0.00$) to $1.62(4)$
($y=1.25$), to be compared with $1.69(4)$ for uncorrelated disorder.
The hyperscaling relation $\gamma/\nu=d-2\beta/\nu$ is therefore not
satisfied {(see values in Tab~\ref{Tab1})}.
We shall identify disorder fluctuations as the origin of this
hyperscaling violation, like in the 3D Random-Field Ising Model (RFIM).
Consider the following decomposition:
 \be\bar\chi=L^d\big[\overline{\langle m^2\rangle}
  -\overline{\langle m\rangle}^2\big]
  -L^d\big[\overline{\langle m\rangle^2}
  -\overline{\langle m\rangle}^2\big]. \label{Eq1}\ee
{
As observed on Fig.~\ref{Chi}, the difference of the two terms
(\ref{Eq1}) leaves an average susceptibility which is smaller
by a factor rougly equal to $3$ in the case of uncorrelated disorder
while it is two orders of magnitude smaller for correlated disorder.
}
The first term, when computed separately, displays a power law behaviour
with an exponent $(\gamma/\nu)^*$ incompatible with $\gamma/\nu$
but in agreement with the hyperscaling relation (Tab.~\ref{Tab1}).
The second term of $(\ref{Eq1})$, the so-called disconnected susceptibility,
involves the ratio
    \be R_m={\overline{\langle m\rangle^2}-\overline{\langle m\rangle}^2
    \over \overline{\langle m\rangle}^2},\label{Rm}\ee
which is expected to behave as $R_m\sim R_m(\infty)+{\cal A}L^{-\phi}$,
if magnetisation is not self-averaging~\cite{WisemanDomany}.
We indeed observe that $R_m$ goes to a non-vanishing constant in the
limit $L\rightarrow +\infty$ (Fig.~\ref{RM}).
The second term of Eq. (\ref{Eq1}) therefore behaves as
$L^d\overline{\langle m\rangle}^2\sim L^{d-2\beta/\nu}$, i.e. with an exponent
satisfying the hyperscaling relation. Since the data indicate that
it is also the case for the first term of Eq. (\ref{Eq1}), one can imagine
that, if their amplitudes are equal, the dominant terms will cancel.
To test this hypothesis, we compute the ratio
  \be {\overline{\langle m^2\rangle}-\overline{\langle m\rangle}^2
  \over\overline{\langle m\rangle^2}-\overline{\langle m\rangle}^2}
  \label{RatioChi}.\ee
We observe a plateau at $1.00(4)$ (Fig.~\ref{RM}). We therefore
conclude that the hyperscaling violation is the result of an exact
cancellation of the dominant contributions of the two terms of
Eq. (\ref{Eq1}). The scaling of the average
susceptibility is therefore determined by the first non-vanishing scaling
correction $\bar\chi\sim L^{d-2\beta/\nu-\omega}$ of any of the two terms
of Eq. (\ref{Eq1}) and the hyperscaling violation exponent $\theta$ is
the exponent $\omega$ of this correction. 
Note that this is also the mechanism of hyperscaling violation
invoked in the context of the RFIM~\cite{SchwartzSoffel} where our
exponent $(\gamma/\nu)^*$ is denoted $4-\bar\eta$~\cite{BrayMoore,Fisher}.
In the case of uncorrelated disorder, the ratio (\ref{RatioChi})
goes to a value significantly different from 1 in
the large size limit (Fig.~\ref{RM}). The dominant contribution
of the two terms of (\ref{Eq1}) do not cancel in this case and
therefore hyperscaling is not violated.

\section{Energy sector}
\begin{figure}
\onefigure[width=\figsize]{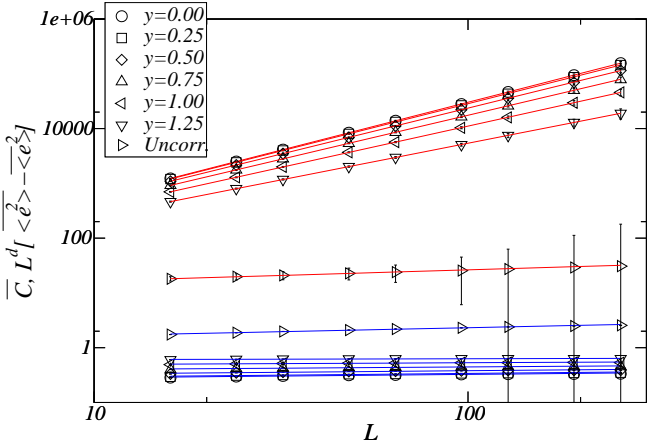}
\caption{Average specific heat $\bar C$ (bottom) and
$L^d[\overline{\langle e^2\rangle}-\overline{\langle e\rangle}^2]$
(top) versus lattice size $L$ for different values of $y$ and
uncorrelated disorder (Uncorr.). The straight lines
correspond to power-law {fits}. For clarity, error bars
of $\bar C$ in the uncorrelated case have been drawn as dashed line
when they overlap with other points.}
\label{C}
\end{figure}

\begin{figure}
\onefigure[width=\figsize]{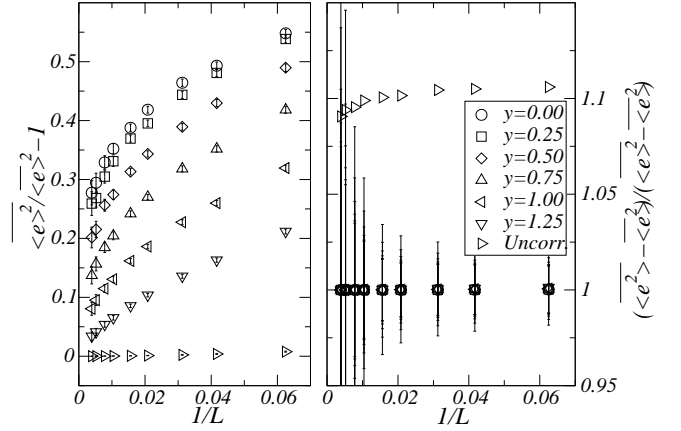}
\caption{{
On the left, ratio defined in the same way as
Eq. (\ref{Rm}), but for energy instead of magnetisation,
versus the inverse of the lattice size $L$. A non-vanishing
asymptotic limit is observed for correlated disorder,
indicating that energy is not self-averaging in this case.
On the right, ratio of the two terms whose difference gives the
average specific heat. Error bars for uncorrelated disorder
are large (they overlap other bars for large lattice sizes)
and have not been represented for clarity. For correlated
disorder, a plateau is observed at a value compatible with 1.}}
\label{Re}
\end{figure}

The divergence of specific heat is completely washed out by
the introduction of disorder, which means that the specific
heat exponent $\alpha/\nu$ is either zero or negative (Fig~\ref{C}).
%The energy on a bond $(i,j)$ is highly correlated to the value
%of the coupling $J_{ij}$. As a consequence, the fluctuations
%$\overline{<e^2>-<e^2>}$ of the total energy,
%and consequently the average specific heat $\bar C$, are small.
The average specific heat
      \be\bar C=L^d\overline{<e^2>-<e>^2}\ee
can be decomposed in the same way as $\bar\chi$:
     \be \bar C=L^d\big[\overline{\langle e^2\rangle}
       -\overline{\langle e\rangle}^2\big]
     -L^d\big[\overline{\langle e\rangle^2}
       -\overline{\langle e\rangle}^2\big].\label{Eq2}\ee
{
Like in the case of susceptibility, the two terms are several
orders of magnitude larger than their difference for
correlated disorder}.
We observe a nice power-law behaviour of the first
term with an exponent in good agreement with $(\alpha/\nu)^*
=(\gamma/\nu)^{\rm AT}=d-a$, which means that the fluctuations of
energy are dominated by the fluctuations of the couplings and
therefore of the polarisation density in the original
Ashkin-Teller model. The second term involves the
ratio $R_e$, constructed in the same way as $R_m$ (\ref{Rm}).
Our numerical data show that energy is not self-averaging
($R_e(\infty)\ne 0$) and the ratio
   \be{\overline{<e^2>}-\overline{<e>}^2
     \over\overline{<e>^2}-\overline{<e>}^2}\ee
 exhibits a plateau at the
value $1.00(5)$ (Fig.~\ref{Re}). This implies the cancellation of
the dominant contribution of the two terms of $\bar C$ so that a violation
of the hyperscaling relation $\alpha/\nu=d-2x_\varepsilon$ is expected.
Even though we cannot measure $x_\varepsilon$ from the scaling
behaviour of energy, we infer that it can be extracted
from the hyperscaling relation $(\alpha/\nu)^*=d-2x_\varepsilon$
which implies $x_\varepsilon=a/2$. In the case of uncorrelated
disorder, $R_e$ is compatible with zero which means that energy is
self-averaging and therefore the two dominant contributions of
Eq. (\ref{Eq2}) do not cancel. As observed, hyperscaling is
not violated in this case.

\begin{figure}
\onefigure[width=\figsize]{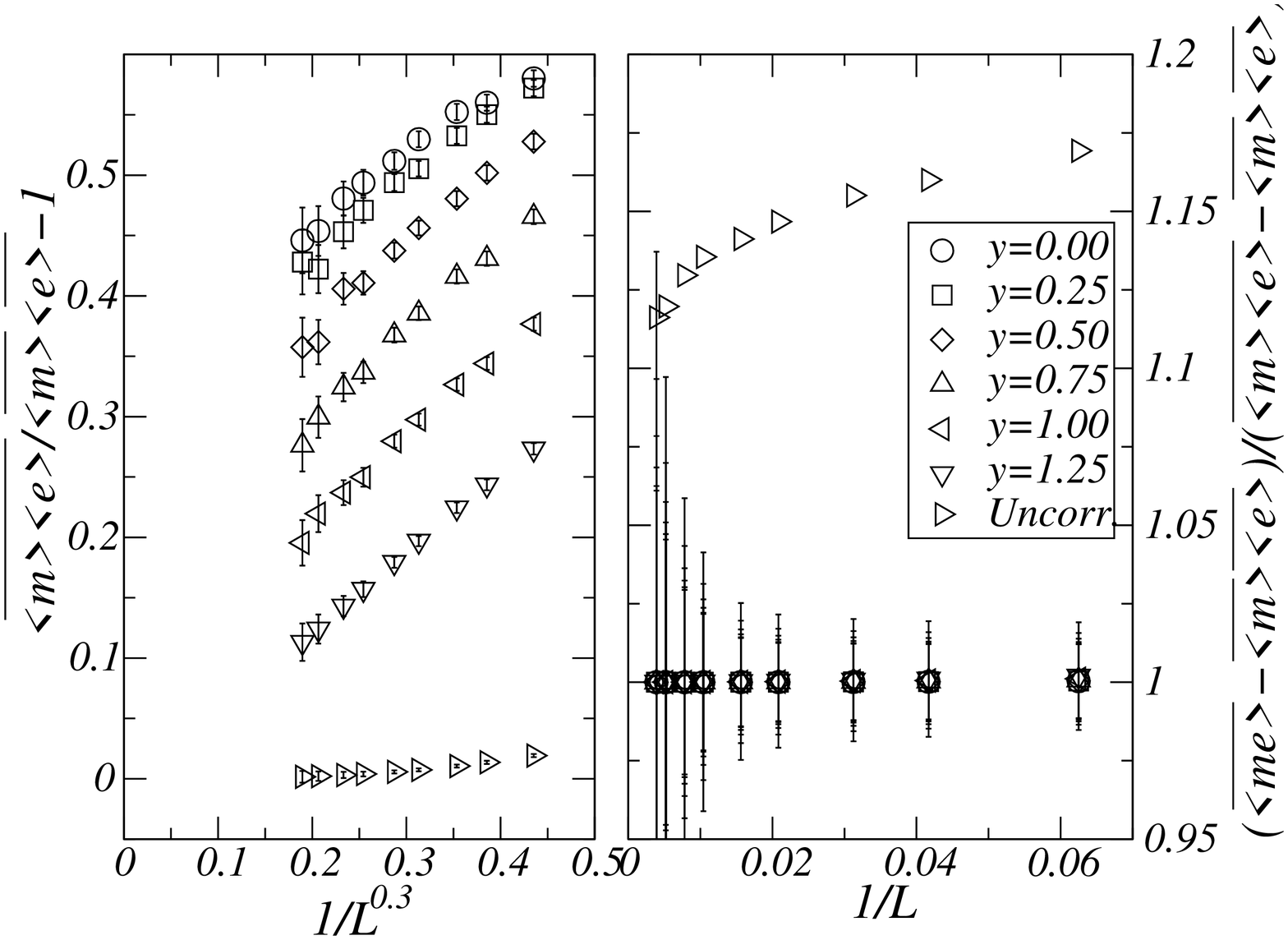}
\caption{{
On the left, ratio defined by Eq. (\ref{Rme}) versus $1/L^{0.3}$.
The non-vanishing extrapolation in the limit $L\rightarrow
+\infty$ indicates that $\langle m\rangle$ and $\langle e\rangle$ are
correlated in the case of correlated disorder. On the right, ratio
(\ref{Rme2}) of the two terms whose difference gives
$-{d\ln\overline{\langle m\rangle}\over d\beta}$. 
Error bars for uncorrelated disorder are large and have not been
represented for clarity. Again, a plateau compatible with $1$ is
observed for correlated disorder.}}
\label{RME}
\end{figure}

In pure systems, a good estimator for the determination of the
correlation {length} exponent $\nu$ is 
  \be-{d\ln\langle m\rangle\over d\beta}
  =L^d{{\langle me\rangle}-{\langle m\rangle\langle e\rangle}\over
  {\langle m\rangle}}.\ee
This is generalised to random systems as:
   \be-{d\ln\overline{\langle m\rangle}\over d\beta}
   =L^d{\overline{\langle me\rangle}
     -\overline{\langle m\rangle\langle e\rangle}
     \over\overline{\langle m\rangle}}.   \label{Eq3}\ee
and is expected to scale as $d-x_\varepsilon$. {A power-law fit
to our data yields} exponents $1/\nu$ close to zero but with large error
bars. Consider again the decomposition
    \be-{d\ln\overline{\langle m\rangle}\over d\beta}
    =L^d{\big[\overline{\langle me\rangle}-\overline{\langle m\rangle}\ 
   \overline{\langle e\rangle}\big]\over\overline{\langle m\rangle}}
   -L^d{\big[\overline{\langle m\rangle\langle e\rangle}
       -\overline{\langle m\rangle}\ \overline{\langle e\rangle}\big]
    \over\overline{\langle m\rangle}}.\label{EqME}\ee
The first term displays a power-law behaviour with exponents
$1/\nu^*$ close to, though slightly above, $d-a/2$, which implies
$x_\varepsilon\simeq a/2$ (Tab.~\ref{Tab1}). This estimate is consistent
with the one obtained from the specific heat. The second term of
(\ref{EqME}) involves the ratio
   \be R_{me}={\overline{\langle m\rangle\langle e\rangle}
   -\overline{\langle m\rangle}\ \overline{\langle e\rangle}
   \over\overline{\langle m\rangle}\ \overline{\langle e\rangle}},
   \label{Rme}\ee
which behaves as $R_{me}(L)\sim R_{me}(\infty)+aL^{-\phi'}$ with
$\phi'\simeq 0.3$. The constant $R_{me}(\infty)$ is clearly finite,
except maybe for $y=1.25$. Like in the magnetic case, the two terms
of Eq. (\ref{EqME}) have the same dominant scaling behaviour.
The ratio
    \be{\overline{\langle me\rangle}-\overline{\langle m\rangle}\ 
    \overline{\langle e\rangle}\over\overline{\langle m\rangle\langle e\rangle}
   -\overline{\langle m\rangle}\ \overline{\langle e\rangle}}\label{Rme2}\ee
displays a plateau at $1.00(4)$ (Fig.~\ref{RME}). Consequently, the
dominant contribution of the two terms of (\ref{EqME}) is the same
and they cancel. Hence, the hyperscaling relation $1/\nu=d-x_\varepsilon$
is expected to be violated. In the case of uncorrelated
disorder, $R_{me}$ is compatible with zero so we do not expect
any cancellation of the two dominant contributions of Eq. (\ref{EqME})
and, consequently, no hyperscaling violation.

\begin{figure}
\onefigure[width=\figsize]{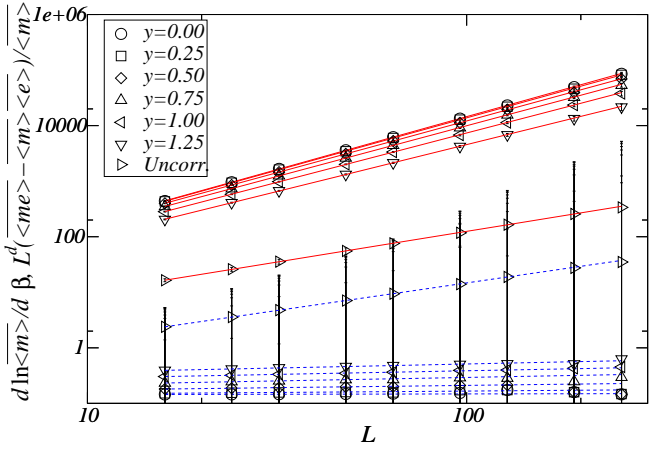}
\caption{Quantity $-{d\ln\overline{\langle m\rangle}\over d\beta}$
(bottom) and $L^d{\overline{\langle me\rangle}
-\overline{\langle m\rangle}\ \overline{\langle e\rangle}
\over\overline{\langle m\rangle}}$ (top) versus lattice size $L$
for different values of $y$ and for uncorrelated disorder (Uncorr.).}
\label{ME}
\end{figure}

\begin{largetable}
\caption{Critical exponents measured by Monte Carlo simulations {(MC)},
or computed from them {(from MC)}, and conjectured values {(Conj.)}.}
\label{Tab1}
\begin{center}
\begin{tabular}{|ll|llllll|}
\hline
&$y$ & $0$ & $0.25$ & $0.5$ & $0.75$ & $1$ & $1.25$
% & Uncorr. Dis.
\\
\hline
&$a$ & $0.25$ & $0.286$ & $0.333$ & $0.4$ & $0.5$ & $0.667$
% & $2$
\\
\hline
$\beta/\nu$ &(MC) & $0.061(5)$ & $0.060(5)$ & $0.067(5)$ &
$0.075(5)$ & $0.091(5)$ & $0.108(4)$
% & $0.150(2)$
\\
$d-2\beta/\nu$ &(from MC) & $1.88(1)$ & $1.88(1)$ & $1.87(1)$ & $1.85(1)$
& $1.82(1)$ & $1.784(8)$
% & $1.700(4)$
\\
$\gamma/\nu$ &(MC) & $1.70(7)$ & $1.69(8)$ & $1.69(7)$ & $1.67(8)$ &
$1.66(6)$ & $1.62(5)$
%& $1.69(5)$
\\
$(\gamma/\nu)^*$ &(MC) & $1.91(2)$ & $1.90(3)$ & $1.89(3)$ &
$1.87(3)$ & $1.83(3)$ & $1.79(3)$
% & $1.76(5)$
\\
\hline
$(\alpha/\nu)^*$ &(MC) & $1.75(1)$ & $1.73(2)$ & $1.68(2)$ & $1.61(2)$ &
$1.51(2)$ & $1.34(2)$
% & $0.20(9)$
\\
$d-a$ &(conj.)& $1.75$ & $1.714$ & $1.667$ & $1.600$ & $1.500$ & $1.333$
% & -
\\
\hline
$1/\nu^*$ &(MC) & $1.90(2)$ & $1.89(2)$ & $1.86(2)$ & $1.83(2)$ & $1.78(2)$
& $1.69(2)$
% & $1.10(7)$
\\
$d-a/2$ &(conj.)& $1.875$ & $1.857$ & $1.835$ & $1.8$ & $1.75$ & $1.667$
% & -
\\
\hline
\end{tabular}
\end{center}
\end{largetable}

\section{Conclusions}
Numerical evidence has been given of the violation of hyperscaling
relations in both the magnetic and energy sectors of the 2D 8-state
Potts model with long-range correlated disorder. Even though this
model is not frustrated, the mechanism causing these violations
was shown to be the same as in the 3D RFIM, namely the cancellation
of the two dominant contributions to the magnetic susceptibility,
specific heat or derivative of the logarithm of magnetisation.
{
However, there are two important differences between our Potts model
with correlated disorder and the RFIM: disorder is coupled to the
energy density and not magnetisation, and the random fixed point
does not lie at zero temperature but at a finite temperature. 
The latter may explain why hyperscaling violation is observed in
both magnetic and energy sectors.} In the magnetic sector,
the hyperscaling violation exponent is estimated to be
$\theta_m=(\gamma/\nu)^*-\gamma/\nu\simeq 0.2$ while in the energy
sector, $\theta_e=(\alpha/\nu)^*-\alpha/\nu\gtrsim d-a$ and
$\theta_e'=(1/\nu)^*-1/\nu\simeq d-a/2$. {
For uncorrelated RFIM, the hyperscaling violation exponent is
compatible with $\theta\simeq d/2$. Since in the absence of
correlation $a=d$, this value is identical to our estimate
$\theta_e'=d-a/2$.}

\acknowledgments
The author gratefully thanks Sreedhar Dutta and the Indian
Institute for Science Education and Research (IISER) of
Thiruvananthapuram for their warm hospitality and a
stimulating environment.

\end{document}